\documentclass{article}

\usepackage{amsthm,amsmath,amssymb}
\usepackage[numbers]{natbib}

\usepackage{hyperref}
\hypersetup{
    colorlinks,%
    citecolor=black,%
    filecolor=black,%
    linkcolor=black,%
    urlcolor=black
}
\usepackage{hypernat}
\usepackage[top=2.5cm, bottom=2.5cm, left=2.8cm,
right=2.8cm]{geometry} 
\usepackage{verbatim}
\usepackage{graphicx}
\usepackage{caption}
\usepackage{subcaption}

\newcommand{\M}{{\mathcal{M}}}

\newcounter{rcnt}[section]

\def\qt#1{\qquad\text{#1}}

\begin{document}

\title{On the impossibility of constructing good population mean
  estimators in a realistic Respondent Driven Sampling model}  

\author{Adityanand Guntuboyina \\ University of California,
  Berkeley\\aditya@stat.berkeley.edu \and Russell Barbour\\ Yale
  University\\ russell.barbour@yale.edu \and Robert Heimer\\ Yale
  University \\robert.heimer@yale.edu}  

\maketitle

\begin{abstract}
Current methods for population mean estimation from data collected by
Respondent Driven Sampling (RDS) are based on the Horvitz-Thompson
estimator together with a set of assumptions on the sampling
model under which the inclusion probabilities can be determined from
the information contained in the data. In this paper, we argue
that such set of assumptions are too simplistic to be realistic and
that under realistic sampling models, the situation is far more
complicated. Specifically, we study a realistic RDS sampling model
that is motivated by a real world RDS dataset. We show that, for this
model, the inclusion probabilities, which are necessary for the
application of the Horvitz-Thompson estimator, can not be determined
by the information in the sample alone. An implication is that, unless
additional information about the underlying population network is
obtained, it is hopeless to conceive of a general theory of population
mean estimation from current RDS data. 
\end{abstract}

%\date{23 September 2010}

%\maketitle

\section{Introduction}
Obtaining useful samples of \textit{hidden} populations \textit{with a
network structure} is a prerequisite for many types of research,
especially for studies of epidemiological problems such as addiction
and HIV/AIDS. Respondent Driven Sampling (RDS) is a recently proposed
sampling technique that seeks to sample from such hidden populations
in a way that allows for valid estimation of population quantities. 

RDS begins with a non-random selection of a  small set of individuals
in the target population. These individuals are referred to as
\textit{seeds}. Data relevant to the study is first collected from
these seeds. In addition, the seeds are asked to report their
degree. We follow Volz and Heckathorn (\citeyear{VolzHeckathorn}) and
define the degree of an individual as the number of people that the
individual could, in principle, recruit. The seeds are then asked, and
provided financial 
incentive, to recruit into the study their social contacts (provided
the contacts are also in the target population). The
sampling continues 
in this way with newly recruited sample members recruiting the next
wave of 
sample members until the desired sample size is reached. Whenever a
subject comes into the study, data relevant to the study are collected
from him/her and in addition, his/her degree in the target population
is recorded. Further, a record of who recruited whom is
maintained.  

While RDS has proved to be extremely effective at penetrating hidden 
populations, the statistical dependencies that it induces in the sample
make the problem of estimating population quantities an intricate
task. Let us now define basic notation to describe the
current employed methods of estimation from RDS data. Let $G$ denote the
population equipped that we wish to sample from. We assume that $G$
has a network/graph structure with nodes/vertices representing and
edges/connections representing extant social relationship. In
particular, the neighbors of an individual denote the set of subjects
that the individual can potentially recruit. Let $y$ denote a
population quantity whose mean, $\mu$, we 
are interested in estimating. A sample of size $n$ is drawn 
via RDS from $G$. Some of the individuals in the sample are
selected as seeds while the others are selected through the process of
recruitment as described above. From each of the sample individuals,
data on the population quantity $y$ is collected. Also, the population
degrees of 
the sample individuals are obtained by enquiry and the information on
recruitment (i.e., who recruited whom) 
is recorded. The goal is to estimate $\mu$ using all available
information i.e., the $y$-values of the sample individuals $y_1,
\dots, y_n$, the population degrees of the sample individuals $d_1,
\dots, d_n$ and the information on who recruited whom.   

The current methods of estimation from RDS data, developed mainly
in Heckathorn (\citeyear{Heckathorn97}), Heckathorn
(\citeyear{Heckathorn02}), Salganik and Heckathorn
(\citeyear{SalganikHeckathorn}), Volz and Heckathorn
(\citeyear{VolzHeckathorn}) and Gile (\citeyear{Gile}) can all be
viewed as being based on the Horvitz-Thompson estimator (Horvitz
and Thompson, \citeyear{HorvitzThompson}), which is a
standard estimator in the theory of survey sampling. The
Horvitz-Thompson estimator for the population mean $\mu$ is given by   
\begin{equation}\label{muht}
  \hat{\mu}_{HT} := \frac{\sum_{i=1}^n y_i/\pi_i}{\sum_{i=1}^n
    1/\pi_i}, 
\end{equation}
where $\pi_i$ is known as the inclusion probability of the $i^{th}$
sample individual and is defined as the probability that the $i^{th}$
sample individual is included in the sample. This
estimator is applicable to both with replacement and without
replacement sampling models. 

It should be noted that the Horvitz-Thompson estimator is applicable
only to probability sampling schemes where the inclusion probabilities
$\pi_1, \dots, \pi_n$ of the sample individuals can be determined. In
the papers cited above, the authors place a variety of assumptions on
the RDS data generation process and argue that, under those
assumptions, the sample obtained according to RDS can be taken to be a
probability sample and that, as required by the Horvitz-Thompson
estimator, that the inclusion probabilities of the
sample individuals can be determined as a function of their
self-reported degrees. We shall describe here the methods of Volz and
Heckathorn (\citeyear{VolzHeckathorn}) and Gile (\citeyear{Gile}). The
estimator in Volz and Heckathorn (\citeyear{VolzHeckathorn}) is termed
RDS II and, as illustrated in Gile and Handcock
(\citeyear{GileHandcockReview}), is an improvement over the classical
RDS estimation procedure developed in Heckathorn
(\citeyear{Heckathorn97}), Heckathorn (\citeyear{Heckathorn02}) and
Salganik and Heckathorn (\citeyear{SalganikHeckathorn}). As a result,
these earlier estimators no longer need to be considered. 

Volz and Heckathorn (\citeyear{VolzHeckathorn}) assume that
respondents recruit uniformly at random from their 
network neighbors (this ensures that RDS is a probability sampling
model). In addition, they also make the following pair of assumptions: 
\begin{enumerate}
\item\label{a1} Samples are drawn \textit{with-replacement} 
\item\label{a2} Each respondent recruits exactly one other respondent
into the study. 
\end{enumerate}
Volz and Heckathorn (\citeyear{VolzHeckathorn}) argue that under these
assumptions, the process of obtaining an RDS sample is equivalent to
performing a random walk on the population network. They then assert,
based on convergence properties of Markov chains, that the inclusion
probability, $\pi_i$, of the $i$th individual in the sample should be
directly proportional to his/her degree, $d_i$. They therefore
construct their estimator for the population mean
$\mu$ by replacing $\pi_i$ in the formula~\eqref{muht} by $d_i$. This
leads to the intuitively appealing and mathematically uncomplicated
estimator RDS II where the population mean is estimated by the
weighted average of the sample observations, the weights being
inversely proportional to the self-reported population degrees of the
sample individuals.  

The problem, however, with RDS II is that it is founded on the two
assumptions~\ref{a1} and~\ref{a2} which are both routinely violated in
practice. Assumption~\ref{a1} which implies that any individual may  
be recruited into the sample more than once, is never
allowed. Assumption~\ref{a2} can be relaxed (see Salganik and  
Heckathorn (\citeyear{SalganikHeckathorn}, pp. 210)) to the case where
all respondents recruit an equal number of respondents. In
practice, however, different respondents recruit differently (we shall
refer to this as differential recruitment in the sequel) and this
aspect is not taken into account by Volz and Heckathorn. More details
on the violation of these assumptions in real-world sampling can be
found in Heimer (\citeyear{Heimer}). 

Regrettably, when assumptions~\ref{a1} and~\ref{a2} are violated, the 
underlying theoretical foundation for  RDS II
breaks down making its role as a population mean estimator unsound and 
suspect. Indeed, if the samples are drawn without replacement, then
the recruitment process is no longer Markov 
because, in standard and near-universal practice, it is not allowed
to re-recruit an individual who is already 
in the sample (however distant past he/she may have been recruited in)
and thus the process is forced to have memory. 

The estimation procedure of Gile (\citeyear{Gile}) is much more
elaborate compared to RDS II. Gile (\citeyear{Gile}) assumed that the
underlying data collection process in RDS can be modeled as a
\textit{successive sampling} process. Under the assumption of
successive sampling, Gile 
described an iterative algorithm for approximating the inclusion
probabilities $\pi_i$ based on the information contained in the RDS
sample alone. The algorithm, which can be considered as a
variant of the Expectation-Maximization (EM) algorithm (see Gile
(\citeyear{Gile}, pp. 12)), further assumes that the population size
is known and also makes an assumption on the graph structure of the
true population (a variant of the configuration model for networks). 

Just like the assumptions of Volz and Heckathorn, the assumption of
successive sampling is also not realistic and would not be a
reasonable approximation to most real-world RDS data collection
processes. For example, although it allows for without-replacement
sampling, it still does not allow for differential
recruitment. Unfortunately, the estimator of Gile (\citeyear{Gile})
crucially depends on 
the assumption of successive sampling and it is not at all clear as to
how it can be extended to real-world situations where the assumptions
of successive sampling do not hold. 

\begin{figure}[htb]
\begin{center}
\includegraphics[scale = 0.7]{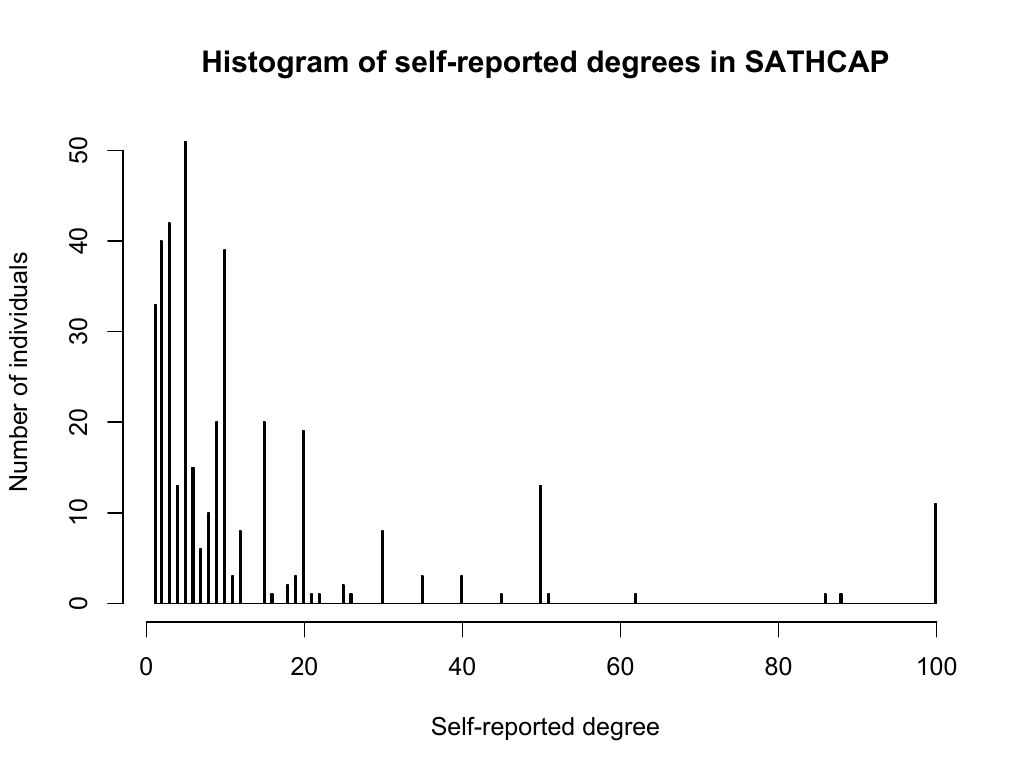}
\caption{Histogram of the self-reported degrees in the SATHCAP
  dataset.}  
\label{hist}
\end{center}
\end{figure}

To illustrate the divergence of the assumptions of Volz and
Heckathorn (\citeyear{VolzHeckathorn}) and those of Gile
(\citeyear{Gile}) with real world RDS sampling, we will review a
recent study of the HIV epidemic in St.Petersburg, Russia. A primary
objective of 
\textit{The Sexual Acquisition and Transmission of HIV Cooperative
  Agreement Project} (SATHCAP) study was to estimate certain
population means (including the prevalence of HIV and hepatitis C) in
the population of Injection Drug  Users (IDUs) in St. Petersburg. In
the first cycle of recruitment conducted from November 2005 through
December 2006, a sample of 373 IDUs was obtained using an RDS
design and data on variables relevant to the study (for example, HIV
status, Hepatitis C status  etc) were collected from the subjects. In
addition, their self-reported population degrees were
recorded. The histogram of these self-reported degrees in the SATHCAP
dataset is given in Figure~\ref{hist}. 

The SATHCAP sample was collected without replacement and
different individuals recruited differently even though each 
individual received the same fixed number of coupons. We have
displayed this differential recruitment pattern in
Table~\ref{obsrec}. It is clear from Table~\ref{obsrec} that the
assumptions of Volz and
Heckathorn (\citeyear{VolzHeckathorn}) and those of Gile
(\citeyear{Gile}) are violated for the SATHCAP dataset. 

\begin{table}
\begin{center}
\begin{tabular}{|c|c|}
\hline
Percentage of Subjects & Number of Recruits \\
\hline
58.98 & 0 \\
15.55 & 1 \\
10.19 & 2 \\
9.65 & 3 \\
3.49 & 4 \\
1.34 & 5 \\
0.8 & 6 \\
\hline
\end{tabular}
\end{center}
\caption{Observed Recruitment Pattern in SATHCAP.  58.98\%
of the sample individuals did not recruit others into the study (in
spite of receiving coupons), 15.55\% of the individuals recruited
exactly one other individual into the study etc.}
\label{obsrec}
\end{table}

The goal of this paper is to argue that for \textit{realistic} RDS
sampling models, it is impossible to determine the inclusion
probabilities $\pi_1, \dots, \pi_n$ from the information available in
an RDS sample. The determination of inclusion probabilities is
possible only under highly simplistic modeling assumptions such as
those made by Volz and Heckathorn (\citeyear{VolzHeckathorn}) or Gile 
(\citeyear{Gile}). 

In section~\ref{model} of this paper, we shall describe a realistic
RDS sampling model, to be denoted by $\M$, that will account for both
differential recruitment and sampling without replacement, thus
creating a more faithful approximation to the data collection 
process in the SATHCAP study. The model $\M$ is conceptually
uncomplicated, although it is more elaborate than currently studied
RDS models since both without replacement sampling and differential
recruitment are assumed. Two
main features of $\M$ are: (a) respondents recruit only from those of
their neighbors who are not already in the sample, and (b) instead of
recruiting exactly one other subject, each respondent recruits no
other subject with a certain probability, exactly one other subject
with a certain probability and so on. We assume that these
probabilities (which will be \textit{parameters} in the model) are the
same for all the respondents. We refer the reader to
section~\ref{model} for a full description of the model.  

In order to use the Horvitz-Thompson estimator for estimating $\mu$
from data generated according to the model $\M$, it is necessary to
determine the inclusion probabilities $\pi_1, \dots, \pi_n$ under
$\M$. The behavior of the inclusion probabilities under $\M$ is quite
detailed involving both individual degrees and the overall network
structure. In section~\ref{sim}, we demonstrate that, in 
contrast to the RDS models studied so far in the literature, the model
$\M$ is such that the inclusion probability for an individual depends
not only on his/her degree but also on the network structure of the
population. We present two simulated population networks having
\textit{identical degrees} but very different inclusion probabilities
under our sampling model. 

Our example demonstrates that the inclusion probabilities $\pi_1,
\dots, \pi_n$ for a realistic RDS model depend not only on the
population degrees of sample individuals, $d_1, \dots, d_n$ but also
on other characteristics of the underlying population
graph. Since the self-reported degrees $d_1, \dots, d_n$ present the
only information about the population network that is contained in an
RDS sample, we deduce that the inclusion probabilities $\pi_1, \dots,
\pi_n$ for realistic RDS models can not be calculated from the RDS
sample. This precludes the application of the Horvitz-Thompson
estimator as a population mean estimator from RDS data. As a result, a
general estimation theory is not possible from RDS data. 

The situation can only be remedied if one obtains additional
information about the underlying population through an RDS sample. We
have some incomplete ideas in this regard that we describe in
Section~\ref{concfu}. We organize concepts as follows:
in the next section, we describe our RDS sampling model $\M$. In
Section~\ref{sim}, we show that inclusion probabilities under $\M$
depend not only on the degrees but also on the structure of the
underlying population network. We finish the paper summarizing our
conclusion and describing some ideas for future work. 

\section{A Realistic Respondent Driven Sampling Model}\label{model}
In this section, we describe a new model for RDS, to be denoted by
$\M$, that allows for both without replacement sampling and
differential recruitment. As explained in the previous section, it is
inspired from the underlying data collection mechanism in the SATHCAP
study. 

Our model $\M$ is conceptually simple. We consider nonnegative real
numbers $p_0, p_1, p_2, \dots$ that sum to 1 and instead of supposing that
every respondent recruits exactly one other subject, we  assume that
each respondent decides to recruit no other subject with probability $p_0$,
exactly one subject with probability $p_1$, exactly two subjects with
probability $p_2$ and so on. This would clearly permit differential
recruitment. Suppose that a respondent decides to recruit $s_1$
subjects and suppose that the number of his/her neighbors that are not
already in the sample is $s_2$. We then assume that the respondent actually
recruits $\min(s_1, s_2)$ subjects \emph{uniformly at random} from among his/her
neighbors who are not already in the sample. As a consequence of this
assumption, respondents recruit only from those of their neighbors
that are not already in the sample. Therefore, $\M$ rules out
subjects reappearing in the sample and hence, is a without-replacement
sampling model.    

We assume that seeds are chosen uniformly at random from all the
population members who are not already in the sample. The other
assumption that is commonly made concerning seed selection in standard
RDS analysis is that
seeds are chosen with probabilities proportional to degrees. It is
debatable as to which assumption (uniformly at random or random with
probabilities proportional to degrees) is more reasonable. We  also 
assume that the sampling starts with one seed and that new seeds are
selected only when necessary i.e., only when recruiting stops. 

The following is the complete description of the sampling model in the
form of a randomized algorithm. The set \textit{active} represents
active/potential recruiters i.e., the sample individuals who currently
possess coupons and who can therefore potentially recruit other
individuals into the sample. When there are 
multiple active recruiters, we assume that one of them (uniformly at
random) recruits first.  
\begin{enumerate}
\item Initially there are no active recruiters. So we initialize
  \textit{active} to be the empty set. The following steps are then
  repeated till the desired sample size $n$ is reached. 
\item If \textit{active} is empty, we need to select a seed. The seed
  is chosen from the set of all population nodes that are \textit{not}
  already in the sample and the seed is included in both the sample
  and the set \textit{active}. 
\item If \textit{active} is non-empty, i.e., if there are active
  recruiters, the following steps take place
\begin{enumerate}
\item One node is chosen uniformly at random from \textit{active},
this respondent recruits first. Let \textit{available} be the set of
all neighbors of this node that are NOT already in the sample and let
$s_1$ be the size of \textit{available}.  
\item A number $s_2$ from $0, 1, \dots$ is chosen with
probabilities $p_0, p_1, \dots$. Then $\min(s_1, s_2)$ nodes are chosen from
\textit{available}. These nodes are included in the sample and in the
set \textit{active}. Also the recruiter is deleted from \textit{active}.
\end{enumerate}
\end{enumerate}
This completes our description of $\M$. For every sampled
individual, we collect data on the study variables and his/her degree
(self-reported). The probabilities $p_0, p_1, \dots$ are parameters in
$\M$. 

In the next section, we demonstrate that for $\M$, the inclusion
probability of an individual depends crucially on the population graph
structure and that the inclusion probabilities can not be determined
by the degrees alone. 
\section{Inclusion probabilities under the model $\M$}\label{sim}
In this section, we argue that, for the sampling model $\M$,
the inclusion probabilities of individuals depend not only on their
degree but also on the network structure of the true population. We
proceed by constructing two population networks $G_1$ and $G_2$ having
the same nodes (say, $N$, of them) and having identical degrees but
vastly different inclusion probabilities under the model $\M$. We
would like to stress that the two networks $G_1$ and $G_2$ have
identical degrees; not just identical degree distributions. In other
words, both $G_1$ and $G_2$ have the same set of nodes, which can be
numbered $1, \dots, N$, and, for each $i = 1, \dots, N$, the degree of
the $i$th vertex is exactly the same in both $G_1$ and $G_2$. 

We base our construction of the networks $G_1$ and $G_2$ on the
degrees reported by the 373 individuals in the SATHCAP sample (see
Figure~\ref{hist}). We take 
these 373 self-reported degrees and resample from this set uniformly
with replacement to create a set of $N$ integers, where $N > 373$ will
be specified shortly. Let us denote this set of $N$ integers by $D_1,
\dots, D_N$ and we assume that $D_1 \geq D_2 \geq \dots \geq D_N$. A
standard result (see, for example, Sierksma and Hoogeveen,
\citeyear{SierksmaHoogeveen}) asserts that there exists a graph 
on $N$ nodes with degrees 
$D_1, \dots, D_N$ (in which case, $D_1, \dots, D_N$ is known as a
\textit{graphical} sequence) provided the following criterion is
satisfied:  
\begin{equation}\label{bollobas}
  \sum_{i = 1}^k \max \left(D_i - k + 1, 0 \right) \leq \sum_{i = k+1}^N D_i
  \qt{for every $k = 1, \dots, N-1$}. 
\end{equation}
Now because the integers $D_1, \dots, D_N$ have been constructed from
the SATHCAP self-reported degrees (which have the histogram given in
Figure 2), the
histogram of $D_1, \dots, D_N$ will also approximately (at least when
$N$ is large) be as in Figure 2 (with just a change of scale on the
$y$-axis). As a result, it is not hard to see 
that, when $N$ is large, the probability that there exists a graph
with degrees $D_1, \dots, D_N$ is quite high (mainly because the right
hand side of~\eqref{bollobas} increases with $N$ while the left hand
side remains unchanged). We choose $N = 5000$ and, at least in
simulations, this invariably resulted in a graphical sequence $D_1,
\dots, D_N$. 

For such a fixed graphical sequence $D_1, \dots, D_N$, we construct
networks $G_1$ and $G_2$ having $N$ nodes with degrees $D_1, \dots,
D_N$. Let us first describe the construction of $G_1$. $G_1$ is a
deterministic graph with degrees $D_1, \dots, D_N$. We take the
algorithm for the construction of $G_1$ from Raman
(\citeyear{Raman}). The algorithm starts with the empty graph (the
graph on $N$ nodes with no edges) and adds edges until the $i$th node
has degree $D_i$ for all $i = 1, \dots, N$. At any stage of the
algorithm, the residual degree 
of the $i$th node is defined as the difference of $D_i$ and its
current degree. The algorithm proceeds by repeatedly joining the node
with the largest residual degree (say, $k$) to the $k$ nodes with the
next largest residual degrees. It can be shown, see Raman
(\citeyear{Raman}), that, whenever $D_1, \dots, D_N$ is graphical,
this algorithm results in a graph with degrees $D_1, \dots, D_N$. From
the construction, it is easy to see that $G_1$ is a deterministic
graph in which the high degree nodes have a tendency to connect to
other high degree nodes. In other words, there is a
\textit{homophily} (the tendency of individuals to associate with
those similar to themselves) by degree in the network $G_1$. 

Let us now explain the construction behind $G_2$ for which we take the
randomized algorithm of Bayati, Kim and Saberi
(\citeyear{BayatiKimSaberi}, pp. 329, Procedure A). This algorithm
also starts with the empty graph and sequentially adds edges between
pairs of non-adjacent nodes until the $i$th node has degree $D_i$ for
all $i = 1, \dots, N$. Unlike Raman's algorithm however, this is a
probabilistic algorithm and at every step, the probability that an
edge is added between two non-adjacent nodes $i$ and $j$ is
proportional to 
\begin{equation*}
  R_i R_j \left(1 - \frac{D_iD_j}{2(D_1 + \dots + D_N)} \right),
\end{equation*}
where $R_i$ is the residual of node $i$. Bayati, Kim and
Saberi (\citeyear{BayatiKimSaberi}) showed that, provided that $D_1$
(which is the maximum of $D_1, \dots, D_N$) is sufficiently small
compared to $D_1 + \dots + D_N$, this algorithm produces a random
graph with degrees $D_1, \dots, D_N$ with high probability and, 
moreover, the distribution of this random graph will be approximately
uniform on the set of all graphs with degrees $D_1, \dots, D_N$. This
condition on $D_1, \dots, D_N$ is satisfied in our case and we can
thus view $G_2$ as one realization of a random graph with degrees
$D_1, \dots, D_N$ whose distribution is approximately uniform over all
graphs with degrees $D_1, \dots, D_N$. 

We have thus created two networks $G_1$ and $G_2$ on $N$ nodes having
identical degrees $D_1, \dots, D_N$. However, these two networks are
very different from each other. $G_1$ is a deterministic graph with a
strong homophily by degree. On the other hand, $G_2$ is one
realization of a random graph that is uniformly distributed over all
graphs with degrees $D_1, \dots, D_N$. We now show that the inclusion
probabilities of the nodes $1, \dots, N$ for a sample of size $n =
373$ (we use $n = 373$ because that was the sample size in the SATHCAP
study) drawn according to the realistic RDS sampling model $\M$ are
vastly different for the two networks $G_1$ and $G_2$. 

\begin{figure}[htb]
\begin{center}
\includegraphics[scale = 0.7]{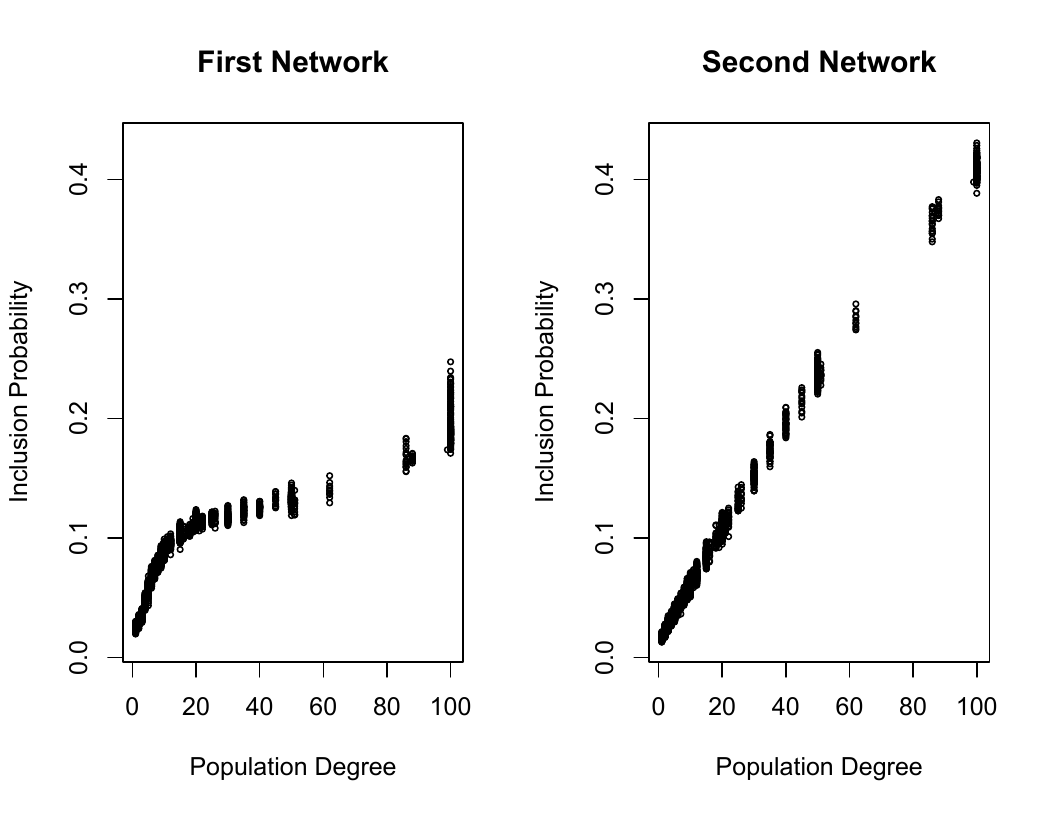}
\caption{For each of the two networks $G_1$ and $G_2$, the inclusion
  probabilities of all population individuals under the model $\M$ are
  plotted against degrees. The scale is exactly the same in both the
  plots.} 
\label{ipplot}
\end{center}
\end{figure}

From each of $G_1$ and $G_2$, we obtained 10000 samples each of 
size $n = 373$ using the model $\M$ with parameters $p_0 = 0.5898$,
$p_1 = 0.1555$, $p_2 = 0.1019$, $p_3=0.0965$, 
$p_4=0.0349$, $p_5=0.0134$, $p_6=0.008$ and $p_7, p_8, \dots$ are all
equal to 0. These specific values for the probabilities were chosen
from the observed recruitment proportions in the SATHCAP
dataset (Table~\ref{obsrec}). Using these 10000 samples, the inclusion
probability of any individual in the population under the model $\M$
(for the chosen parameters) can be well-approximated by the proportion of
samples containing the individual. These inclusion probabilities are
plotted against degrees for each of the two networks $G_1$ and $G_2$
in Figure~\ref{ipplot}. It is clear from Figure~\ref{ipplot} that the
inclusion probabilities for $G_1$ and $G_2$ are quite and meaninfully
distinct (the 
scale of the two plots is exactly the same) even though $G_1$ and
$G_2$ have identical degrees. The inclusion probabilities for
individuals in $G_1$ are roughly proportional to the square root of
their degrees. On the other hand, the inclusion probabilities in $G_1$
are nearly directly proportional to degrees. Therefore, the inclusion
probabilities for the model $\M$ depend not only the degrees but also
on the network structure of the true population.  

\section{Conclusion and Future Work}\label{concfu}
In contrast with earlier works on respondent driven
sampling, we presented a realistic RDS model, motivated by
a real world RDS dataset. We demonstrated that for this model, the
inclusion probabilities of sample individuals does not depend on their
degrees alone. Indeed, we constructed two population networks $G_1$
and $G_2$ having identical set of nodes and degrees but with vastly
different inclusion probabilities. This implies that, for this RDS 
model, it is impossible to construct a general estimation theory for
population means based on the Horvitz-Thompson estimator.   

For estimation under such a realistic RDS model to be feasible, we
either need more information about the underlying population network
or we need to be content in only estimating certain but not all
population means. In this regard, we have the following ideas which we
hope to explore in future work: 
\begin{enumerate}
\item Under assumptions in Volz and Heckathorn
  (\citeyear{VolzHeckathorn}), the inclusion probabilities of sampled
  individuals are proportional to their degrees and hence they depend
  solely on the degrees up to the constant of
  proportionality. Heckathorn (\citeyear{Heckathorn97}) realized 
  that the researcher can learn the degrees of individuals by just
  asking them and since then, sampled individuals are always asked to
  report their degrees in RDS studies. In the case of the model $\M$
  however, we have seen that the inclusion probabilities depend not
  only on the degrees but also on the network structure of the true
  population. It is of interest to understand the precise network
  characteristics on which the inclusion probabilities depend and
  whether such characteristics can be learned by asking the sampled
  subjects additional questions. Recent work by Cepeda et al
  (\citeyear{Cepeda2011}) suggests that additional information on
  network structure such as total network size and stability would
  enhance HIV prevalence estimates and assist prevention measures in a
  population of injection drug users. These types of questions could
  yield enough information about the underlying network for
  the purpose of approximating inclusion probabilities. 
\item We have studied two network models in this paper; the ones that
  were involved in the construction of $G_1$ and $G_2$. It will be of
  interest to study more such network models and to investigate the
  behavior of inclusion probabilities under them.  
\item It will be of interest to explore alternative RDS estimators
  that are not based on the Horvitz-Thompson estimator. Such
  estimators may not have the general applicability of the
  Horvitz-Thompson estimator but may work in certain special
  instances. For example, suppose that the population quantity whose
  mean we are interested in estimating is distributed as a Bernoulli
  random variable with probability of success equal to $p$ and suppose
  that $p$ is independent of all features of the network. In that
  case, it should be clear that the sample mean is as good as any
  other estimator. In other words, if the quantity of interest does
  not depend on the network features, then the sample mean is an
  adequate estimator. Drawing from this intuition, it is reasonable to
  believe that for population quantities that do not depend
  significantly on the degrees, one might not need to use the
  Horvitz-Thompson estimator as simpler estimates based perhaps on the 
  sample mean might be adequate. Making such an idea rigorous will
  require more work. 
\end{enumerate}

\bibliographystyle{plainnat}
\bibliography{AG}

\def\noopsort#1{}
\begin{thebibliography}{12}
\providecommand{\natexlab}[1]{#1}
\providecommand{\url}[1]{\texttt{#1}}
\expandafter\ifx\csname urlstyle\endcsname\relax
  \providecommand{\doi}[1]{doi: #1}\else
  \providecommand{\doi}{doi: \begingroup \urlstyle{rm}\Url}\fi

\bibitem[Bayati et~al.(2007)Bayati, Kim, and Saberi]{BayatiKimSaberi}
Mohsen Bayati, Jeong~Han Kim, and Amin Saberi.
\newblock A {S}equential {A}lgorithm for {G}enerating {R}andom {G}raphs.
\newblock In \emph{Approximation, Randomization, and Combinatorial
  Optimization}, volume 4627 of \emph{Lecture Notes in Computer Science}, pages
  326--340, Berlin/Heidelberg, 2007. Springer.

\bibitem[Cepeda et~al.(2011)Cepeda, Odinokova, Heimer, Grau, Lyubimova,
  Safiullina, Levina, and Niccolai]{Cepeda2011}
Javier~A Cepeda, Veronika~A Odinokova, Robert Heimer, Lauretta~E Grau,
  Alexandra Lyubimova, Liliya Safiullina, Olga~S Levina, and Linda~M Niccolai.
\newblock Drug network characteristics and hiv risk among injection drug users
  in russia: the roles of trust, size, and stability.
\newblock \emph{AIDS and Behavior}, 15\penalty0 (5):\penalty0 1003--1010, 2011.

\bibitem[Gile(2010)]{Gile}
Krista~J. Gile.
\newblock Improved inference for respondent-driven sampling data with
  application to {HIV} prevalence estimation.
\newblock \textit{Under review}, available at http://arxiv.org/abs/1006.4837,
  2010.

\bibitem[Gile and Handcock(2010)]{GileHandcockReview}
Krista~J. Gile and Mark~S. Handcock.
\newblock Respondent-driven sampling: {A}n assessment of current methodology.
\newblock \emph{Sociological Methodology}, page forthcoming, 2010.

\bibitem[Heckathorn(1997)]{Heckathorn97}
Douglas~D. Heckathorn.
\newblock Respondent-driven sampling: A new approach to the study of hidden
  populations.
\newblock \emph{Social Problems}, 44:\penalty0 174--199, 1997.

\bibitem[Heckathorn(2002)]{Heckathorn02}
Douglas~D. Heckathorn.
\newblock Respondent-driven sampling {II}: Deriving valid population estimates
  from chain-referral samples of hidden populations.
\newblock \emph{Social Problems}, 49:\penalty0 11--34, 2002.

\bibitem[Heimer(2005)]{Heimer}
Robert Heimer.
\newblock Critical issues and further questions about respondent-driven
  sampling: comment on ramirez-valles, et al.(2005).
\newblock \emph{AIDS and Behavior}, 9\penalty0 (4):\penalty0 403--408, 2005.

\bibitem[Horvitz and Thompson(1952)]{HorvitzThompson}
D.~G. Horvitz and D.~J. Thompson.
\newblock A generalization of sampling without replacement from a finite
  universe.
\newblock \emph{Journal of the American Statistical Association}, 47:\penalty0
  663--685, 1952.

\bibitem[Raman(1991)]{Raman}
Rajeev Raman.
\newblock Generating random graphs efficiently.
\newblock In \emph{Advances in Computing and Information - ICCI '91}, volume
  497 of \emph{Lecture Notes in Computer Science}, pages 149--160,
  Berlin/Heidelberg, 1991. Springer.

\bibitem[Salganik and Heckathorn(2004)]{SalganikHeckathorn}
Matthew~J. Salganik and Douglas~D. Heckathorn.
\newblock Sampling and estimation in hidden populations using respondent-driven
  sampling.
\newblock \emph{Sociological Methodology}, 34:\penalty0 193--239, 2004.

\bibitem[Sierksma and Hoogeveen(1991)]{SierksmaHoogeveen}
Gerard Sierksma and Han Hoogeveen.
\newblock Seven criteria for integer sequences being graphic.
\newblock \emph{Journal of Graph theory}, 15\penalty0 (2):\penalty0 223--231,
  1991.

\bibitem[Volz and Heckathorn(2008)]{VolzHeckathorn}
Erik Volz and Douglas~D. Heckathorn.
\newblock Probability based estimation theory for respondent-driven sampling.
\newblock \emph{Journal of Official Statistics}, 24:\penalty0 79--97, 2008.

\end{thebibliography}

\end{document}